\documentclass[aps,pra,reprint,superscriptaddress]{revtex4-2}

\usepackage[T1]{fontenc}
\usepackage{lmodern}
\usepackage{amsmath,amssymb,bm}
\usepackage{physics}
\usepackage{bbm}
\usepackage{times}
\usepackage{graphicx}
\usepackage[colorlinks=true,linkcolor=blue,citecolor=blue,urlcolor=blue]{hyperref}

\newcommand{\Hw}{\mathcal H_w}
\newcommand{\Hc}{\mathcal H_c}
\newcommand{\sgn}{\mathrm{sgn}}
\newcommand{\Nneg}{\mathcal{N}}

\begin{document}

\title{Operationally Admissible Post-Quantum Correlations from a Standard Quantum Walk}

\author{Marcos C.\ de Oliveira}
\affiliation{Instituto de F\'{\i}sica Gleb Wataghin, Universidade Estadual de Campinas, Campinas, SP, Brazil}

\date{\today}

\begin{abstract}
It is shown that a standard one-dimensional coined discrete-time quantum walk can generate
operationally admissible post-quantum correlations in a coin--position Bell scenario, without
any modification of its unitary nearest-neighbor dynamics.
Post-quantum features enter exclusively through an extended operational preparation of the coin,
described by a complementarity-violating Hermitian trace-one operator
$\rho_c(\mathbf r)=\frac12(\mathbb I+\mathbf r\cdot\boldsymbol\sigma)$ with $\|\mathbf r\|>1$,
while physical consistency is enforced solely at the level of observable statistics via
admissibility and no-signaling.
The extended preparation admits an experimental emulation through a two-component quasiprobability
reconstruction over physical coin states, at the price of an increased sampling overhead.
The walk-generated coin--position entanglement can support CHSH values exceeding
Tsirelson’s bound, even though the walk dynamics remains fully standard. We also show that physically natural coarse-grained position measurements can render such post-quantum
correlations operationally inaccessible, strongly suppressing observable Bell violations. 
The purpose here is to contrast the separation between the
existence of post-quantum behavior and its accessibility under realistic measurement constraints.
\end{abstract}

\maketitle

\section{Introduction}

Bell nonlocality provides one of the sharpest operational signatures of nonclassical correlations.
In the standard bipartite scenario with two dichotomic measurements per party, the
Clauser--Horne--Shimony--Holt (CHSH) inequality \cite{CHSH1969} bounds the strength of correlations
compatible with local hidden-variable models, while quantum theory admits violations up to
Tsirelson’s bound \cite{Tsirelson1980}.
At an abstract level, it is known that correlations can exceed Tsirelson’s bound while still
respecting no-signaling \cite{PopescuRohrlich1994}.
Such post-quantum correlations play a central role in generalized probabilistic theories and in
information-theoretic reconstructions of quantum theory
\cite{Barrett2005,Brunner2014, PhysRevLett.101.020401}.
Related operational viewpoints connect post-quantum behaviors to quasiprobability representations and negativity as an
operational resource, as well as to modern discussions of contextuality/nonclassicality beyond state positivity
(see, e.g., Refs.~\cite{PhysRevLett.101.020401,Ferrie2011,Veitch2012,Barrett2007GPT,Pashayan2015}).
Identifying concrete and physically transparent dynamical mechanisms that generate operationally consistent
post-quantum correlations---rather than merely postulating them abstractly---remains a major open challenge.

Beyond their formal existence, a key issue is the \emph{operational accessibility} of nonlocal
correlations under realistic measurement constraints.
In complex or high-dimensional systems, experimentally natural measurement families often involve
coarse-graining, limited resolution, or restricted control over relevant subspaces.
In this context, the absence of a Bell violation in a given test does not necessarily imply the
absence of nonclassical correlations in the underlying state, but may instead reflect the limited
ability of available measurements to access them.
This motivates a controlled setting in which one can disentangle two logically independent
questions: (\textit{i}) whether a given dynamical process can support post-quantum correlations
\emph{in principle}, and (\textit{ii}) whether such correlations remain detectable under physically
natural measurement restrictions.

For two parties holding qubits, quantum correlations generated from positive bipartite states
$\rho_{AB}\ge 0$ and local dichotomic observables necessarily obey Tsirelson’s bound,
$|S|\le2\sqrt2$~\cite{Tsirelson1980}.
Therefore, any operationally consistent attempt to obtain $|S|>2\sqrt2$ must move beyond the
standard quantum state space and/or measurement postulates.
In our approach this is achieved by relaxing preparation positivity while enforcing physical
consistency only at the level of the observed probabilities through admissibility and
no-signaling.
However, within a strictly qubit--qubit scenario such an extension would be operationally
unilluminating -- it would not naturally separate the state-space extension from a controlled
dynamical mechanism that generates experimentally meaningful measurement structure. To extend to larger Hilbert spaces comprising a qubit-qudit system would be the minimal possible choice. In that sense,
quantum walks provide precisely the missing ingredients, as it  involves a minimal two-level coin
coupled to a high-dimensional walker subsystem (qudit) through fully standard, local unitary dynamics,  with the advantage of being experimentally implementable.
This realizes a qubit--qudit Bell scenario in which admissible post-quantum correlations could be engineered, in principle \cite{deOliveira2020Complementarity}, but its accessibility is yet  to be comproved.

In this work we address these questions within the minimal and well-controlled framework of a
\emph{standard} coined discrete-time quantum walk (DTQW).
Discrete-time quantum walks provide a paradigmatic model for coherent quantum dynamics on a lattice,
with applications ranging from quantum algorithms and transport to quantum simulation of condensed
matter and relativistic phenomena
\cite{Aharonov1993,Kempe2003,Venegas-Andraca2012,portugal2013quantum,wang2013physical}.
Their appeal stems from strict locality at the level of the step operator, interference generated
by repeated unitary evolution, and analytical and numerical tractability.
Importantly for our purposes, DTQWs naturally generate entanglement between a two-level coin and a
rapidly expanding walker-position Hilbert space, thereby offering a tunable and conceptually simple
route to high-dimensional bipartite correlations under fully local, unitary dynamics.
Crucially, the walk dynamics itself is left entirely unchanged -- the step operator is unitary,
nearest-neighbor, and identical to that used in conventional quantum walks.
The departure from standard quantum correlations arises exclusively from the coin preparation,
which we treat in an extended operational framework based on complementarity violation.
Specifically, we consider Hermitian trace-one coin operators of the Bloch form
$\rho_c(\mathbf r)=\frac12(\mathbb I+\mathbf r\cdot\boldsymbol\sigma)$ with $\|\mathbf r\|>1$,
which are nonpositive and therefore fall outside the quantum state space.
Rather than imposing positivity as a primitive axiom, we enforce physical consistency solely at
the level of observed statistics -- all joint probabilities entering the Bell test must be
nonnegative and normalized (admissibility) and satisfy no-signaling.
Operational frameworks in which complementarity---rather than positivity---is taken as primitive
have been shown to naturally give rise to post-quantum features while preserving no-signaling at
the level of observable statistics \cite{deOliveira2020Complementarity}.
The present work provides a concrete realization of this perspective within the controlled
dynamics of a standard coined quantum walk. It is important to remark that $\rho_c(\mathbf r)$ with $\|\mathbf r\|>1$ is not a physical quantum state and is not
interpreted as something prepared in a single run. Rather, it serves as an operational device to
define target joint statistics, which can be emulated experimentally through a quasiprobability
reconstruction over physical coin preparations, while admissibility and no-signaling are enforced
directly at the level of observed probabilities.

A natural concern is whether restricting attention to admissible parameter choices amounts to a form of
postselection.
It does not--- no conditioning on measurement outcomes is performed, and no runs are discarded.
Rather, admissibility and no-signaling are 
\emph{defining operational consistency conditions} for the reconstructed
probabilities produced by the extended preparation rule.
Outside the admissible region the reconstruction does not define a valid operational experiment, because the
resulting numbers cannot be interpreted as joint probabilities.
Throughout, we report only those parameter regimes in which the reconstructed statistics define a bona fide
non-signaling behavior.

\begin{figure*}[t!]
  \centering
  \includegraphics[width=\textwidth]{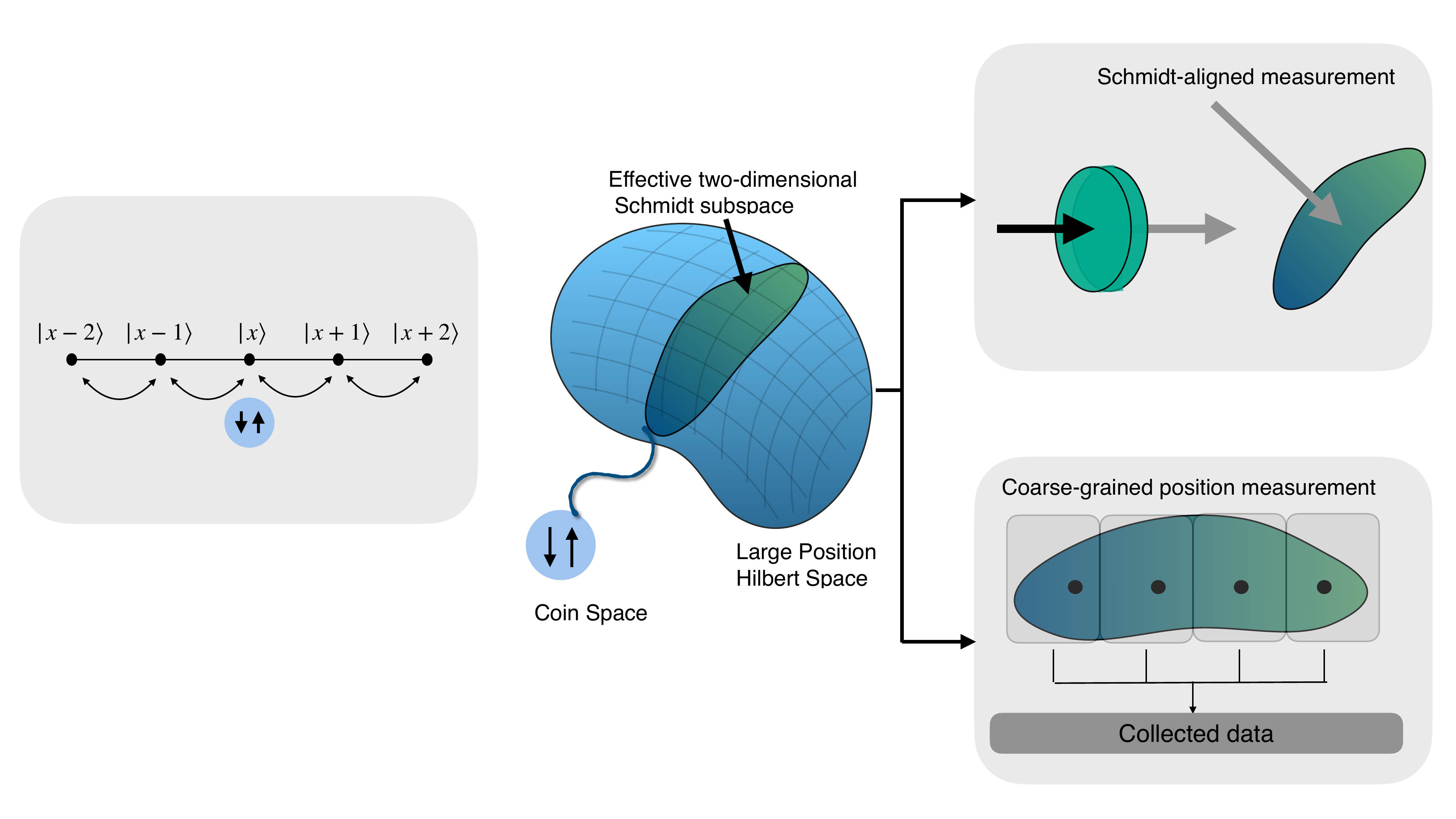}
  \caption{Schematic illustration of the separation between the existence and operational accessibility of post-quantum correlations in a standard coined quantum walk.
  Left: a fully standard, unitary nearest-neighbor quantum walk on a one-dimensional lattice, with a two-level coin controlling the conditional shift.
  Center: entanglement between the coin (two-dimensional Hilbert space) and the walker position generates an effective low-dimensional Schmidt subspace within the large position Hilbert space.
  Right: two measurement regimes. Schmidt-aligned observables access this subspace and reveal admissible post-quantum CHSH violations, while coarse-grained position measurements fail to resolve it, strongly suppressing observable Bell correlations.}
\label{fig:schematic}
\end{figure*}

Our results establish a sharp separation between the \emph{existence} and \emph{accessibility} of
post-quantum correlations in this setting.
On the one hand, we construct Schmidt-aligned Bell measurements that optimally probe the effective
two-dimensional Schmidt subspace of the final coin--position operator, thereby revealing
operationally admissible CHSH violations beyond Tsirelson’s bound.
On the other hand, we show that physically natural coarse-grained position measurements can fail
to resolve this subspace, strongly suppressing observable Bell violations even when
post-quantum correlations exist in principle.
This separation is schematically summarized in Fig.~\ref{fig:schematic}.
Therefore: 
(i) For standard Hadamard-walk dynamics and extended coin preparations with $\|\mathbf r\|>1$, there exist
measurement choices for which the reconstructed behavior $p(a,b|i,j)$ is normalized, nonnegative, and non-signaling,
while the CHSH value exceeds Tsirelson’s bound;
(ii) The accessibility of such behavior depends strongly on measurement structure -- Schmidt-aligned observables can
reveal admissible post-quantum values, while physically natural diagonal-in-position coarse-grainings can suppress
observable violations at large walk times, despite a finite-time window where violations remain detectable.
The present results position quantum walks as a controlled operational platform for probing
the boundary between quantum and post-quantum correlations.
They highlight the nontrivial role of measurement structure in witnessing post-quantum behavior in
large Hilbert spaces, and provide an explicit example in which standard local unitary dynamics can
support operationally admissible post-quantum correlations under extended preparations.

This paper is organized as follows.
Section~II introduces the standard coined quantum walk and the extended operational preparation
framework.
In Sec.~III we define the coin--position Bell scenario and the admissibility and no-signaling
conditions imposed on the observed statistics.
Section~IV presents the numerical results, highlighting the separation between existence and
operational accessibility of post-quantum correlations.
Section~V concludes with a discussion and outlook.


\section{Background}

\subsection{Coined discrete-time quantum walks}
A coined discrete-time quantum walk on $\mathbb Z$ is defined on the tensor product
$\Hw=\ell^2(\mathbb Z)$ and $\Hc=\mathbb C^2$.
Each step consists of a unitary coin operation acting on $\Hc$, followed by a conditional shift on
$\Hw$ that moves the walker to neighboring lattice sites depending on the coin state.
Interference arises from the repeated application of these operations, while locality is enforced
by the strictly nearest-neighbor structure of the shift.
This construction is entirely standard and serves as a model of coherent, local quantum
dynamics on a lattice.
Standard reviews can be found in Refs.~\cite{Kempe2003,Venegas-Andraca2012}.

We consider the standard one-dimensional coined discrete-time quantum walk with step operator
\begin{equation}
U = S\,(\mathbb I_w\otimes H),
\end{equation}
where $H$ is the Hadamard coin,
\begin{equation}
H=\frac{1}{\sqrt2}
\begin{pmatrix}
1&1\\
1&-1
\end{pmatrix},
\end{equation}
and $S$ is the conditional nearest-neighbor shift,
\begin{equation}
S=\sum_{x\in\mathbb Z}
\Big(\ket{x-1}\!\bra{x}\otimes\ket{0}\!\bra{0}
+\ket{x+1}\!\bra{x}\otimes\ket{1}\!\bra{1}\Big).
\end{equation}

\subsection{Extended preparations and complementarity violation}
In the extended preparation framework considered here, the coin is initialized in a Hermitian,
trace-one operator of the Bloch form
\begin{equation}
\rho_c(\mathbf r)=\frac12\big(\mathbb I+\mathbf r\cdot\boldsymbol\sigma\big),
\qquad \mathbf r\in\mathbb R^3,
\end{equation}
which corresponds to a physical density operator only when $\|\mathbf r\|\le1$.
For $\|\mathbf r\|>1$, $\rho_c(\mathbf r)$ is not positive and therefore does not represent a
physical quantum state.
Rather than interpreting such operators as states, we treat them as elements of an extended
operational preparation framework.
From an operational perspective, this corresponds to a violation of complementarity at the level of
preparation, a setting shown to naturally give rise to post-quantum features while preserving
no-signaling at the level of observable statistics~\cite{deOliveira2020Complementarity}.

In this framework, positivity is not imposed as a primitive axiom.
Instead, physical consistency is enforced solely at the level of observable statistics -- all joint
outcome probabilities accessible in a given measurement scenario must be nonnegative, normalized,
and satisfy no-signaling.
Violations of positivity at the preparation level therefore manifest only indirectly, through the
structure of admissible correlations, and may enable post-quantum behavior while remaining
operationally consistent.
This stance aligns naturally with generalized probabilistic and operational approaches to
nonclassical correlations
\cite{PopescuRohrlich1994,Barrett2005,Brunner2014}.

Although $\rho_c(\mathbf r)$ is not positive for $\|\mathbf r\|>1$ and therefore does not represent
a physical quantum state, it is a well-defined operator within the extended operational framework
introduced above.
Concretely, any such operator admits a signed decomposition into physical coin states,
$\rho_c(\mathbf r)=w_+\,\rho_{+\hat{\mathbf r}}+w_-\,\rho_{-\hat{\mathbf r}}$, with
$w_\pm=(1\pm\|\mathbf r\|)/2$.
All observable quantities entering the Bell test can therefore be reconstructed by performing
standard quantum-walk experiments with the physical coin states $\rho_{\pm\hat{\mathbf r}}$
and combining the resulting statistics with classical signed weights.
Operational consistency is enforced not at the level of preparation, but at the level of
observable joint probabilities, which are explicitly required to be normalized, nonnegative,
and no-signaling.
Within this framework, post-quantum correlations arise as admissible operational correlations,
while the underlying walk dynamics and measurements remain fully standard and experimentally
accessible.

The initial joint operator is
\begin{equation}
\rho_0=\rho_w(0)\otimes\rho_c(\mathbf r),
\end{equation}
with,
\begin{equation}
\rho_w(0)=\ket{0}\!\bra{0}.
\end{equation}
Its evolution under the standard walk dynamics is given by unitary conjugation,
\begin{equation}
\rho_T=U^T\rho_0(U^\dagger)^T,
\end{equation}
which preserves Hermiticity and trace for all $T$.
Any physical constraints therefore arise exclusively at the level of observable measurement
statistics, which are explicitly enforced in the Bell scenario considered below.

\section{Coin--position Bell scenario}

We consider a bipartite Bell scenario in which Alice measures the coin degree of freedom and Bob
measures the walker position after $T$ steps of the quantum walk.
Although the walker Hilbert space is infinite-dimensional, the Bell scenario is well defined because
Bob’s observables are dichotomic.
As a result, standard CHSH bounds apply independently of the underlying Hilbert-space dimension.
It is important to emphasize that Bob does not perform a full position-resolving measurement in the
Bell test. Instead, the position outcome $x$ is mapped into a binary variable
$b=B_j(x)\in\{\pm1\}$, corresponding to a yes/no question about the walker’s location.
Operationally, this defines a dichotomic observable with eigenvalues $\pm1$, irrespective of the
infinite dimensionality of the underlying position Hilbert space.
The Bell scenario is therefore fully specified by two binary measurement settings per party, and
standard CHSH bounds apply without modification.

In the present work, however, we consider two conceptually distinct strategies for Bob’s
measurement, which are essential to interpret our results. In the first strategy, Bob is allowed to
implement optimized dichotomic observables acting effectively on the two-dimensional Schmidt
subspace of the walker–coin state. These measurements are not constrained to be diagonal in the
position basis and correspond to ideal projective measurements that maximize the Bell violation.
This strategy establishes the \emph{existence} of nonclassical correlations in the generated state.
In the second strategy, Bob is restricted to physically motivated measurements obtained from
position detection followed by classical coarse-graining, as described below. These observables
are diagonal in the position basis and correspond to experimentally accessible procedures. The
contrast between these two strategies highlights the difference between the existence of
nonclassical correlations in the state and their accessibility under realistic measurement
constraints.

\subsection{Measurements}

\paragraph{Alice (coin).}
Alice performs two dichotomic measurements on the coin,
\begin{equation}
A_0=\hat{\mathbf a}\cdot\boldsymbol\sigma,
\qquad
A_1=\hat{\mathbf a}'\cdot\boldsymbol\sigma,
\end{equation}
with $\|\hat{\mathbf a}\|=\|\hat{\mathbf a}'\|=1$.

\paragraph{Bob (walker position).}
We consider two distinct classes of dichotomic measurements on the walker.

\emph{(i) Schmidt-aligned observables (existence).}
In the idealized scenario, Bob performs optimized dichotomic measurements acting on the
effective two-dimensional Schmidt subspace of the walker–coin state. For the pure state
$\rho_T=\ket{\Psi_T}\!\bra{\Psi_T}$, we consider its Schmidt decomposition with respect to the
coin--position bipartition,
\begin{equation}
\ket{\Psi_T} = \sqrt{\lambda_0}\,\ket{0}_c \ket{\phi_0}_w
+ \sqrt{\lambda_1}\,\ket{1}_c \ket{\phi_1}_w,
\label{schmidt}\end{equation}
where $\{\ket{\phi_0},\ket{\phi_1}\}$ are orthonormal states in the walker Hilbert space and
$\lambda_0+\lambda_1=1$.

The relevant dynamics is therefore confined to the two-dimensional subspace
$\mathcal{H}_{\mathrm{eff}}=\mathrm{span}\{\ket{\phi_0},\ket{\phi_1}\}$.
Within this subspace, Bob’s observables can be represented as effective Pauli operators,
\begin{eqnarray}
\tilde{\sigma}_x &=& \ket{\phi_0}\!\bra{\phi_1} + \ket{\phi_1}\!\bra{\phi_0},\\ 
\tilde{\sigma}_y &=& -i\ket{\phi_0}\!\bra{\phi_1} + i\ket{\phi_1}\!\bra{\phi_0}, \\
\tilde{\sigma}_z &=& \ket{\phi_0}\!\bra{\phi_0} - \ket{\phi_1}\!\bra{\phi_1}.
\end{eqnarray}

Bob’s dichotomic measurements are then given by
\begin{equation}
B_j^{(\mathrm{opt})} = \hat{\mathbf b}_j \cdot \tilde{\boldsymbol\sigma},
\end{equation}
with $\|\hat{\mathbf b}_j\|=1$.
These observables are not diagonal in the position basis and correspond to ideal projective
measurements that maximize the CHSH value according to the Horodecki criterion.
This strategy provides an upper bound on the Bell violation compatible with the generated state
and certifies the existence of nonclassical correlations.

\emph{(ii) Coarse-grained position observables (accessibility).}
In the physically constrained scenario, Bob measures the walker position and applies a
deterministic binning $B_j(x)\in\{\pm1\}$. We consider, in particular, the sign binning
$B_0(x)=\sgn(x)$ and a single-threshold binning $B_1(x)=+1$ for $|x|\ge x_0$ and $-1$
otherwise. The corresponding observables are diagonal in the position basis and can be written as
\begin{equation}
B_j=\sum_{x\in\mathbb Z} B_j(x)\ket{x}\!\bra{x}.
\end{equation}
These measurements define the experimentally accessible Bell scenario studied in this work.

\subsection{Correlators and CHSH value}
For each pair of settings $(i,j)$, the correlators are defined as
\begin{equation}
E_{ij}=\Tr\big[(B_j\otimes A_i)\rho_T\big],
\end{equation}
from which the CHSH parameter is constructed as
\begin{equation}
S=E_{00}+E_{01}+E_{10}-E_{11}.
\end{equation}

\subsection{Observed probabilities, admissibility, and no-signaling}
Given settings $(i,j)$ and outcomes $a,b=\pm1$, the observed joint probabilities are
\begin{equation}
p(a,b|i,j)=\Tr\!\left[\left(\Pi^{(j)}_{b}\otimes \Pi^{(i)}_{a}\right)\rho_T\right],
\end{equation}
where $\Pi^{(i)}_{a}=\frac12(\mathbb I+aA_i)$ and
$\Pi^{(j)}_{b}=\frac12(\mathbb I+bB_j)$ are the projectors associated with the dichotomic observables.

A parameter choice is said to be \emph{admissible} if
$p(a,b|i,j)\ge0$ for all $a,b,i,j$ and
$\sum_{a,b}p(a,b|i,j)=1$ for each pair $(i,j)$.
In addition, we explicitly enforce \emph{no-signaling} at the level of the observed probabilities
\begin{align}
\sum_{b=\pm1} p(a,b|i,j) &= \sum_{b=\pm1} p(a,b|i,j'), \quad \forall\, a,i,\, j,j', \\
\sum_{a=\pm1} p(a,b|i,j) &= \sum_{a=\pm1} p(a,b|i',j), \quad \forall\, b,j,\, i,i'.
\end{align}
When these conditions hold, the reconstructed tables define a valid non-signaling behavior in the
standard (2,2,2) CHSH scenario.

For $\|\mathbf r\|\le1$, corresponding to positive coin preparations, the observed correlations are
bounded by Tsirelson’s inequality, $|S|\le2\sqrt2$.
For $\|\mathbf r\|>1$, admissible and no-signaling correlations exceeding Tsirelson’s bound may
occur.

\section{Results}
We now present our numerical results, organized around the two complementary measurement scenarios
that address logically distinct questions.
First, we apply the Schmidt-aligned benchmark Bell test that characterizes the maximal
correlations supported by the walk-generated coin--position state.
Second, we restrict Bob to simple  coarse-grained position measurements in order to probe the
operational accessibility of post-quantum correlations under physically natural measurement
constraints.

\subsection{Schmidt-aligned benchmark (existence of post-quantum correlations)}
We begin by establishing whether the coin--position state generated by the walk is, in principle,
capable of supporting post-quantum correlations.
To this end, we construct a benchmark Bell test in which the measurement observables are aligned
with the Schmidt decomposition of the final coin--position state.
This benchmark simultaneously serves as a stringent validation of the Bell-test implementation
and as a characterization of the maximal Bell correlations supported by the state.

For $\|\mathbf r\|=1$, the evolution produces a pure bipartite state
$\ket{\psi_T}\in\Hc\otimes\Hw$ with exactly two nonzero Schmidt coefficients.
In this case, the maximal quantum violation of the CHSH inequality is determined solely by these
coefficients via the Horodecki criterion \cite{Horodecki1995}.
For the time-step $T=60$, we obtain Schmidt coefficients $\{0.843,\,0.538\}$, corresponding to a theoretical
quantum maximum $S_{\max}=2.70$.
By explicitly constructing the associated optimal observables and embedding them into the full
coin and position Hilbert spaces, we recover this value exactly in our numerical implementation.
This agreement provides a stringent validation of both the walk dynamics and the Bell-test
construction.

Having established this benchmark, we fix the Schmidt-aligned observables and extend the coin
preparation beyond the quantum domain to $\|\mathbf r\|>1$, while enforcing admissibility and
no-signaling at the level of the joint outcome probabilities.
For a range of preparation magnitudes and directions, we find admissible violations exceeding
Tsirelson’s bound.
A representative witness at $T=60$ achieves
\begin{equation}
|S| = 3.0810,
\qquad
\min_{a,b,i,j}p(a,b|i,j)=3.27\times10^{-3},
\end{equation}
for $\|\mathbf r\|=1.45$.

\begin{figure}[t]
\centering
\includegraphics[width=\linewidth]{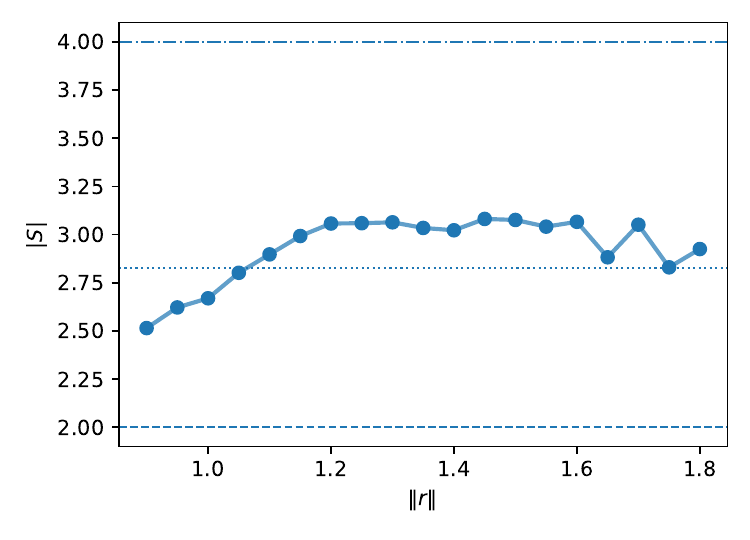}
\caption{CHSH value $|S|$ versus extended Bloch magnitude $\|\mathbf r\|$ for the fixed Schmidt-aligned benchmark at $T=60$.
Markers show the discrete maximizers obtained from the directional scan at each sampled $\|\mathbf r\|$, and the solid curve is a guide to the eye connecting these points.
Horizontal reference lines indicate the classical CHSH bound $|S|=2$ (dashed), Tsirelson’s bound $|S|=2\sqrt2$ (dotted), and the algebraic maximum $|S|=4$ (dash-dotted).}
\label{fig:B_S_vs_r}
\end{figure}

\begin{figure}[t]
\centering
\includegraphics[width=\linewidth]{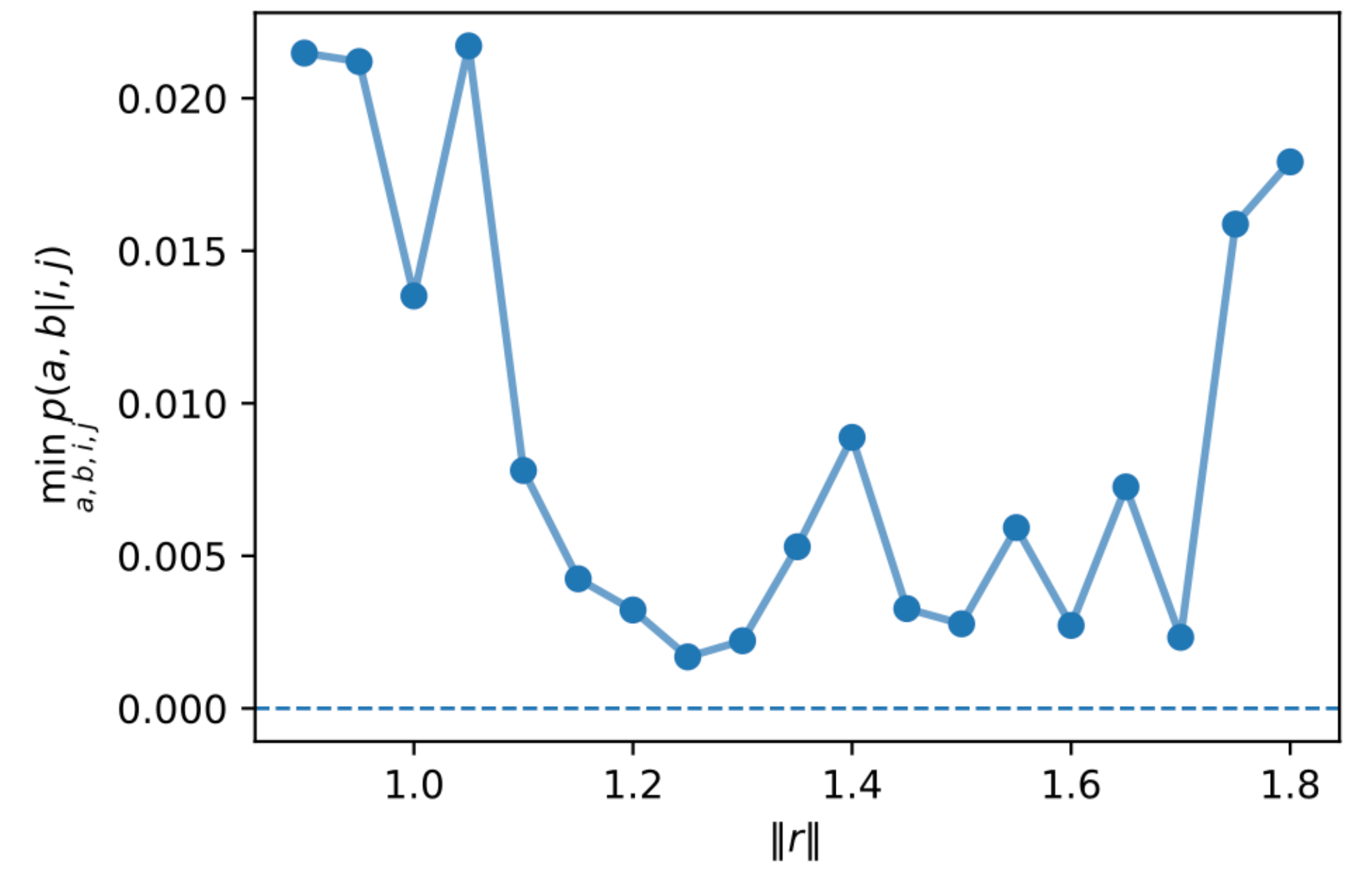}
\caption{Minimum joint probability $\min_{a,b,i,j} p(a,b|i,j)$ for the Schmidt-aligned benchmark,
demonstrating admissibility of the post-quantum violations. The dashed horizontal line indicates the boundary
$\min p = 0$.}
\label{fig:B_minp_vs_r}
\end{figure}

Figure~\ref{fig:B_S_vs_r} shows the resulting CHSH value as a function of $\|\mathbf r\|$ for this
fixed measurement choice, while Fig.~\ref{fig:B_minp_vs_r} displays the corresponding admissibility
margin.
Together, these results demonstrate that a completely standard coined quantum walk can support
admissible post-quantum correlations at the level of appropriately chosen measurements,
independently of any coarse-graining of the walker position.
We have verified that the qualitative behavior reported here persists over a broad range of step
numbers $T$, with $T=60$ chosen as a representative value for which the Schmidt structure and the
effects of coarse-graining discussed below are both clearly resolved.

\begin{figure}[t]
\centering
\includegraphics[width=\linewidth]{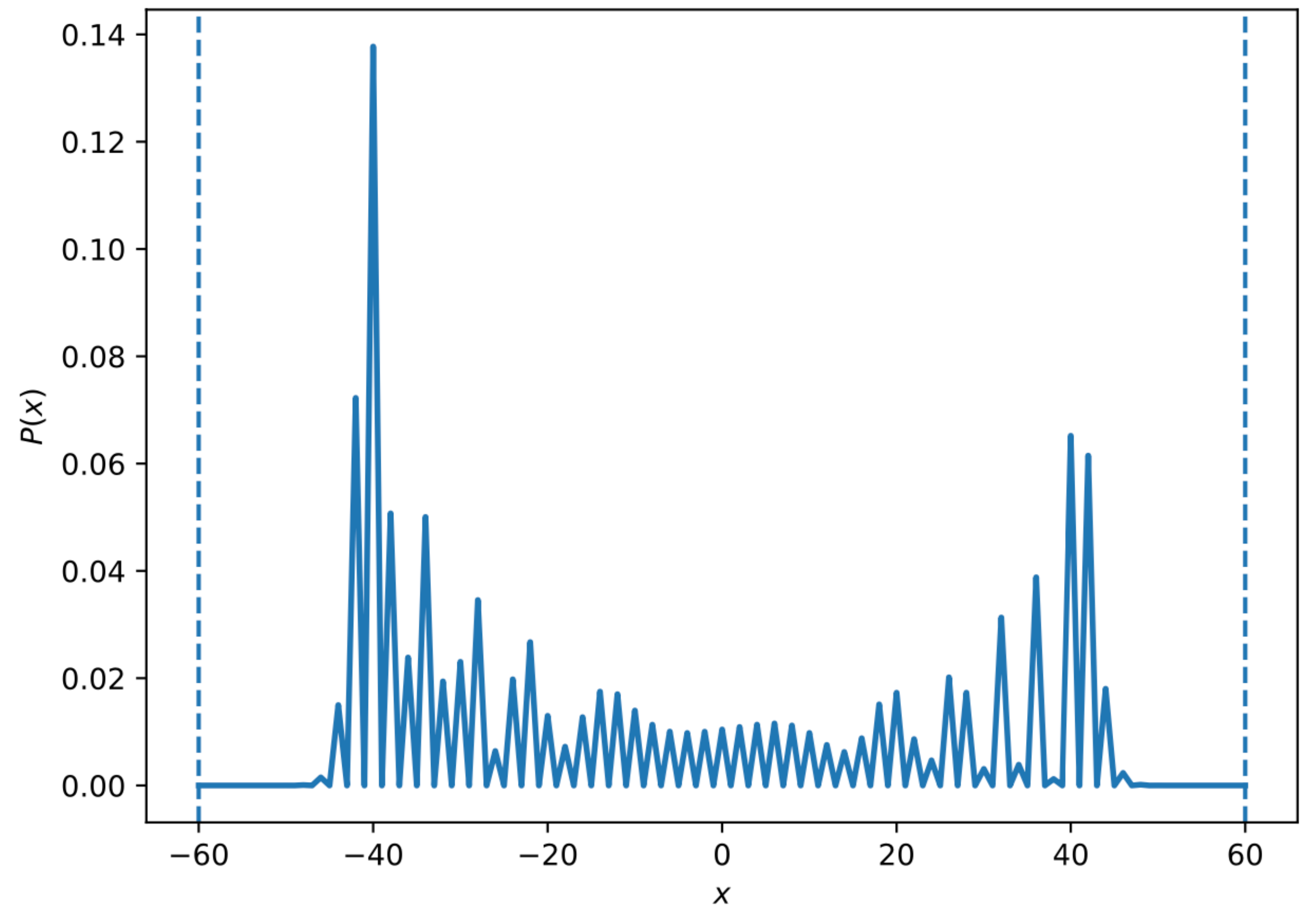}
\caption{Final-time walker position distribution $P(x)$ at $T=60$ for the maximal-violation
Schmidt-aligned witness at $\|\mathbf r\|=1.45$, with $|S|=3.0810$.
The distribution remains confined to $|x|\le T$ and exhibits the familiar interference-broadened
profile of a standard Hadamard coined quantum walk.}
\label{fig:B_Px}
\end{figure}

To complement the Bell-test data, Fig.~\ref{fig:B_Px} shows the final-time position distribution
$P(x)$ of the walker for the representative Schmidt-aligned witness with $|S|=3.0810$ at
$T=60$ and $\|\mathbf r\|=1.45$.
This confirms that the walk kinematics is unchanged (standard light cone and Hadamard-type ballistic profile);
the observed post-quantum CHSH values originate solely in the extended preparation together with the Bell-test choice.

It is important to emphasize that the Schmidt-aligned Bell test accesses correlations that live in
a highly structured two-dimensional \emph{mode subspace} of the walker Hilbert space, rather than in
a simple coarse partition of position space.
At fixed $T$ and for a positive preparation ($\|\mathbf r\|=1$), the walk generates a pure bipartite
state $\ket{\psi_T}\in\Hc\otimes\Hw$ with Schmidt rank at most two,
\begin{equation}
\ket{\psi_T}
=
s_0\,\ket{u_0}_c\otimes\ket{b_0}_w
+
s_1\,\ket{u_1}_c\otimes\ket{b_1}_w,
\end{equation}
where $\{\ket{u_0},\ket{u_1}\}$ and $\{\ket{b_0},\ket{b_1}\}$ are orthonormal Schmidt bases for coin
and walker, respectively, and $s_0,s_1>0$ are Schmidt coefficients, simmilar to Eq. (\ref{schmidt}).
The corresponding optimal CHSH observables for Bob act as Pauli operators in the Schmidt subspace
$\mathrm{span}\{\ket{b_0},\ket{b_1}\}$ and can be embedded into the full position Hilbert space as
\begin{equation}
B_j
=
\sum_{\mu,\nu=0}^1 (B_j^{\mathrm{sub}})_{\mu\nu}\,\ket{b_\mu}\!\bra{b_\nu}
+(\mathbb I-P),\;
P=\sum_{\mu=0}^1\ket{b_\mu}\!\bra{b_\mu},
\label{eq:Schmidt_embed_B}
\end{equation}
so that $\pm1$ outcomes are assigned trivially outside the Schmidt support.
Crucially, the Schmidt vectors $\ket{b_\mu}$ are generally delocalized superpositions over many
lattice sites,
$\ket{b_\mu}=\sum_x \beta_x^{(\mu)}\ket{x}$, so that the operators $B_j$ contain large
off-diagonal coherences in the position basis, through terms such as
$\ket{b_0}\!\bra{b_1}+\ket{b_1}\!\bra{b_0}$.
While Schmidt-aligned measurements provide a ceiling on the post-quantum correlations supported by the
walk-generated coin--position state, implementing them experimentally would require coherent mode mixing
(interferometric measurements) across many walker positions, which is experimentally demanding in generic quantum-walk platforms. This motivates, in the following sections,
an operationally natural restriction in which Bob is limited to coarse-grained position measurements, allowing
us to quantify the accessibility of post-quantum correlations under realistic measurement constraints.

\subsection{Coarse-grained position measurements (operational accessibility)}

We now turn to the question of operational accessibility.
Specifically, we restrict Bob to simple and physically natural coarse-grained position
measurements, while allowing Alice to optimize over two coin observables.
This setting is not designed to maximize Bell violations, but rather to probe how measurement
restrictions affect the observability of post-quantum correlations.

We consider the sign binning $B_0(x)=\sgn(x)$ together with a single-threshold binning
$B_1(x)=+1$ for $|x|\ge x_0$ and $-1$ otherwise.
An initial coarse scan over thresholds indicates that the best admissible witnesses for $T=60$
and $\|\mathbf r\|$ in the range of interest cluster near $x_0/T\simeq0.6$.
Motivated by this observation, we fix $x_0/T=0.6$ and refine the optimization over (i) the
extended preparation direction $\hat{\mathbf r}$ at fixed magnitude $\|\mathbf r\|$, and (ii) the
two coin measurement directions $\hat{\mathbf a}$ and $\hat{\mathbf a}'$.
The coarse-grained optimization is performed via a randomized multi-start search over the preparation direction
$\hat{\mathbf r}$ (at fixed $\|\mathbf r\|$) and Alice’s measurement directions $(\hat{\mathbf a},\hat{\mathbf a}')$,
while Bob’s binning is fixed as specified above.
At each trial we compute the reconstructed joint tables $p(a,b|i,j)$ and retain only those candidates that satisfy
admissibility and no-signaling within numerical tolerance; details are given in Appendix~\ref{sec:numerics}.

This procedure explicitly separates the two roles played by the extended preparation:
it allows $\rho_c(\mathbf r)$ to be nonpositive for $\|\mathbf r\|>1$, while ensuring that the only
operationally accessible quantities entering the Bell test---the observed joint probabilities---remain
valid and no-signaling.

For $T=60$ and $\|\mathbf r\|=1.45$, the best admissible witness found within this restricted
measurement family yields
\begin{equation}
|S| \approx 1.6665 < 2,
\end{equation}
with admissibility margin
$\min_{a,b,i,j}p(a,b|i,j)\approx2.18\times10^{-2}$.
Representative parameters are summarized in Table~\ref{tab:A_best}.
Although this does not violate the classical CHSH bound, it is informative precisely because it
occurs in the same regime in which the Schmidt-aligned benchmark demonstrates that post-quantum
correlations exist for suitable measurements.

\begin{table}[t]
\caption{Representative best witness obtained under coarse-grained position measurements
at $T=60$ and $\|\mathbf r\|=1.45$, with $B_0(x)=\sgn(x)$ and a single-threshold binning
$B_1(x)=+1$ for $|x|\ge x_0$ and $-1$ otherwise.}
\label{tab:A_best}
\begin{ruledtabular}
\begin{tabular}{lc}
Quantity & Value \\
\hline
$T$ & $60$ \\
$\|\mathbf r\|$ & $1.45$ \\
$x_0/T$ & $0.6$ \\
$|S|$ & $1.6665$ \\
$\min_{a,b,i,j}p(a,b|i,j)$ & $2.18\times10^{-2}$ \\
$(E_{00},E_{01},E_{10},E_{11})$ & $(0.6324,\,0.2198,\,0.6627,\,-0.1517)$ \\
\end{tabular}
\end{ruledtabular}
\end{table}

\begin{figure}[t]
\centering
\includegraphics[width=\linewidth]{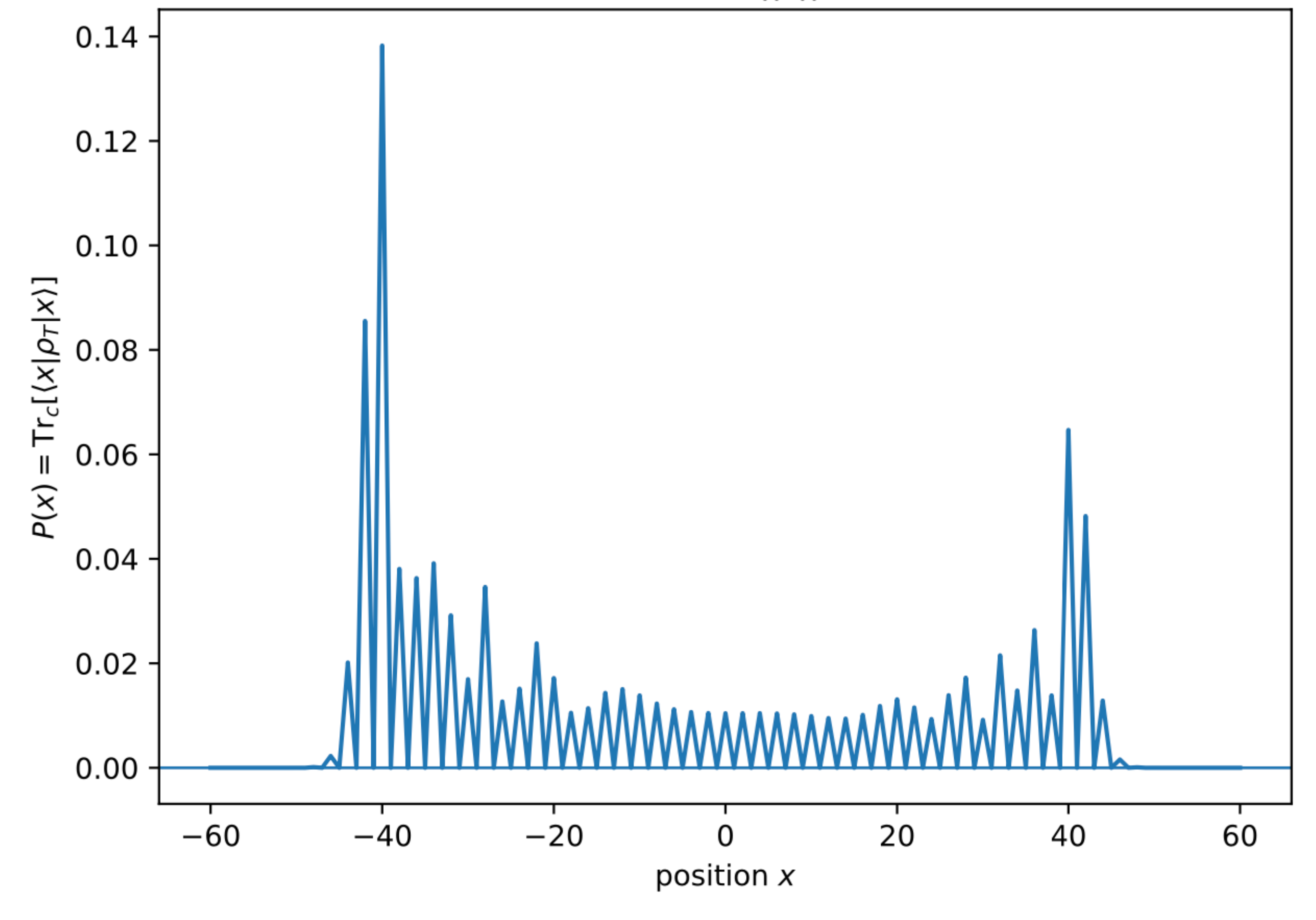}
\caption{Final-time walker position distribution $P(x)$ at $T=60$ for the coarse-grained witness of
Table~\ref{tab:A_best}. The distribution remains confined to $|x|\le T$ and shows the familiar
interference-broadened profile of a standard Hadamard coined quantum walk.}
\label{fig:A_Px}
\end{figure}

To complement the Bell-test data, Fig.~\ref{fig:A_Px} shows the final-time position distribution
$P(x)$ of the walker for the coarse-grained witness of Table~\ref{tab:A_best}, obtained from the
position marginal of $\rho_T$.
This plot serves two purposes.
First, it emphasizes that the walk dynamics remains fully standard: the distribution is confined
to the usual light cone $|x|\le T$ and exhibits the familiar interference-broadened profile of a
Hadamard walk.
Second, it illustrates that the strong suppression of Bell correlations is not due to
any anomalous transport feature, but rather to measurement restrictions: the coarse binning does
not effectively access the two-dimensional Schmidt subspace that governs the maximal Bell
correlations revealed in the benchmark test.

These results highlight a sharp contrast with the Schmidt-aligned benchmark.
Although the walk-generated coin--position state is capable of supporting large post-quantum
correlations, simple coarse-grained position measurements fail to access the effective
two-dimensional Schmidt subspace responsible for the strongest violations.
As a consequence, observable Bell correlations can be strongly suppressed, even when
post-quantum correlations exist in principle.
This separation is not an artifact of modified dynamics: the step operator and locality structure
remain fully unchanged across both scenarios.

\subsection{Finite-time accessibility of post-quantum correlations under coarse graining}
\label{subsec:finiteT}

The analysis above focused on a fixed walk time $T=60$, for which coarse-grained position
measurements fail to access the post-quantum correlations revealed by the Schmidt-aligned benchmark.
However, the suppression observed at large $T$ raises a natural question:
is coarse-graining intrinsically incompatible with post-quantum correlations,
or is the observed inaccessibility a consequence of Hilbert-space growth
outpacing measurement resolution?

To address this question, we perform a systematic finite-time analysis of the coarse-grained operational search, keeping the
coarse-grained measurement structure fixed while varying the number of steps $T$.
Specifically, we fix the extended preparation magnitude to $\|\mathbf r\|=1.45$ and repeat the
randomized optimization over preparation directions and coin observables for a sequence of small
walk times $T\in\{2,4,6,8,10\}$.
For each $T$, Bob’s measurements are restricted to the same sign binning $B_0(x)=\sgn(x)$ and a
single-threshold binning $B_1(x)$, with the threshold ratio $x_0/T$ optimized over a discrete grid.
All candidates are required to satisfy admissibility and no-signaling, as in the large-$T$ analysis.

\begin{figure}[t]
\centering
\includegraphics[width=\linewidth]{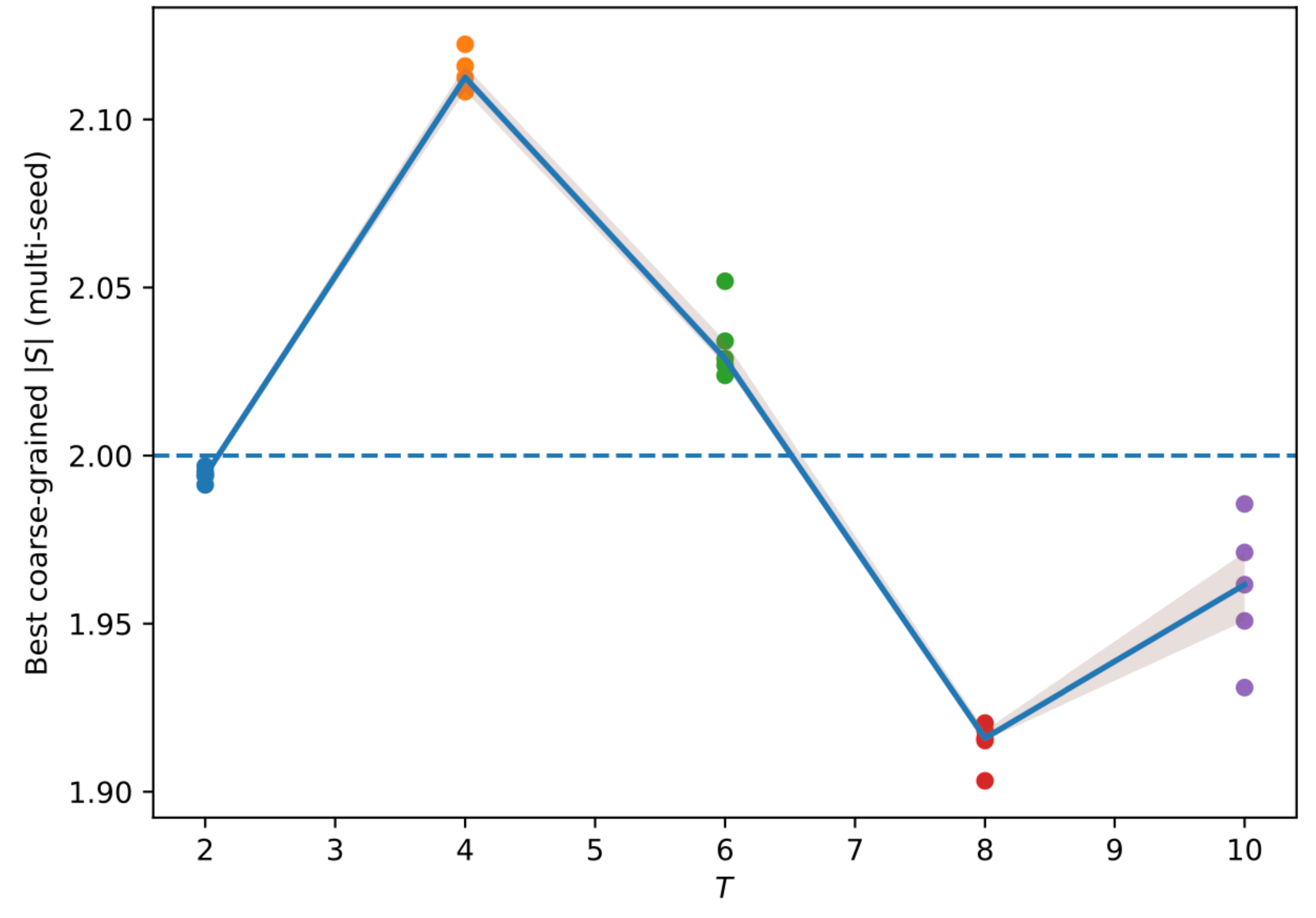}
\caption{Finite-time coarse-grained CHSH performance for $\|\mathbf r\|=1.45$.
For each $T\in\{2,4,6,8,10\}$ we perform a randomized search over preparation directions
$\hat{\mathbf r}$ and coin observables $(A_0,A_1)$, while Bob is restricted to the coarse-grained
position measurements described in the main text and the threshold ratio $x_0/T$ is selected from
a fixed grid. Markers show the best admissible value found for each random seed, and the solid curve
indicates the median over seeds. A clear finite-time window ($T=4$ and $T=6$) exhibits robust violations
of the classical CHSH bound.}
\label{fig:A_S_vs_T}
\end{figure}

\begin{figure}[t]
\centering
\includegraphics[width=\linewidth]{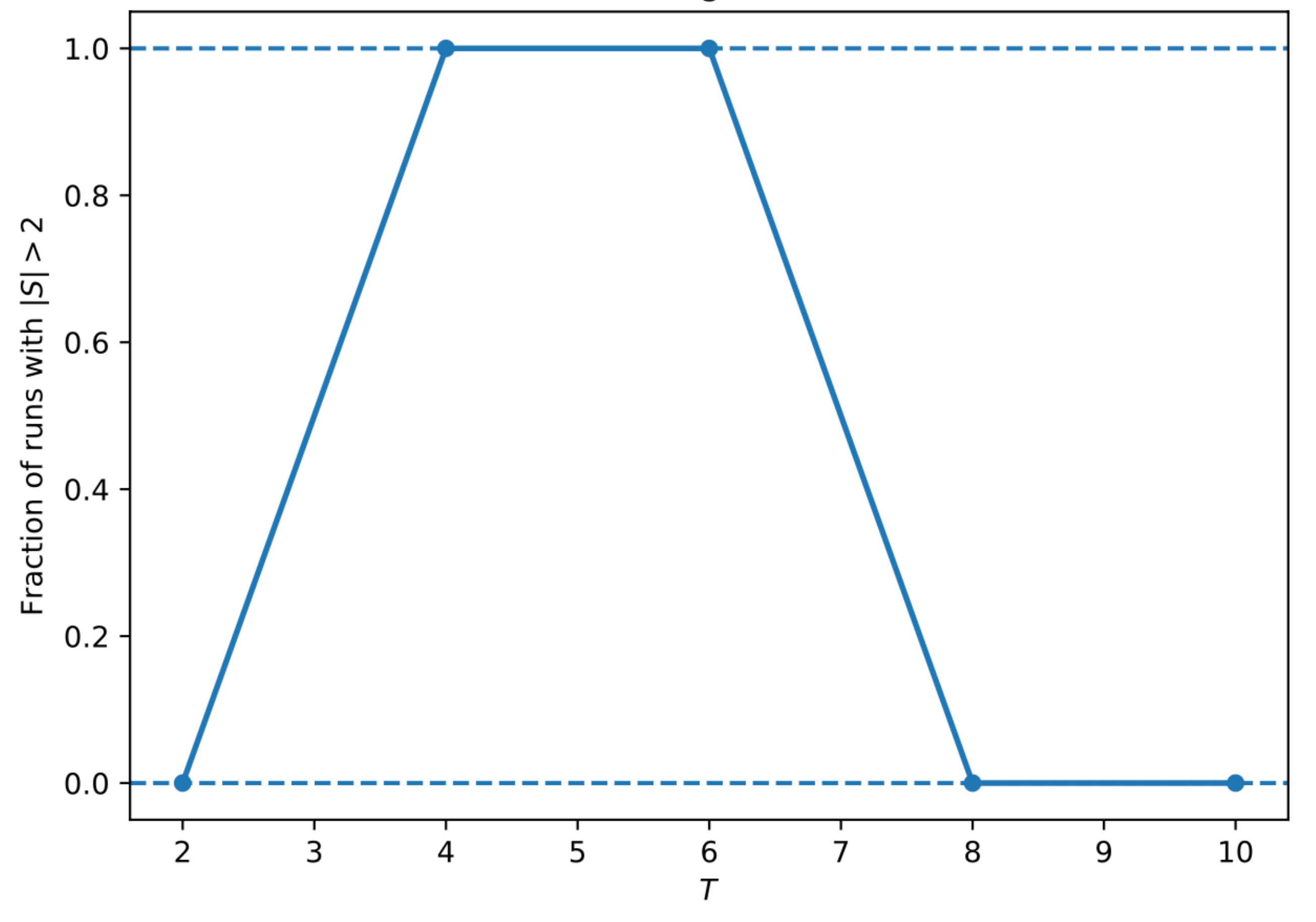}
\caption{Robustness of finite-time coarse-grained violations.
For each $T$, we report the fraction of accepted admissible candidates (over repeated randomized trials and multiple seeds)
that yield $|S|>2$ under the coarse-grained position measurements.
Violations are most readily accessible at small $T$, and become strongly suppressed as the position Hilbert space expands.}
\label{fig:A_frac_gt2}
\end{figure}

\begin{figure}[t]
\centering
\includegraphics[width=\linewidth]{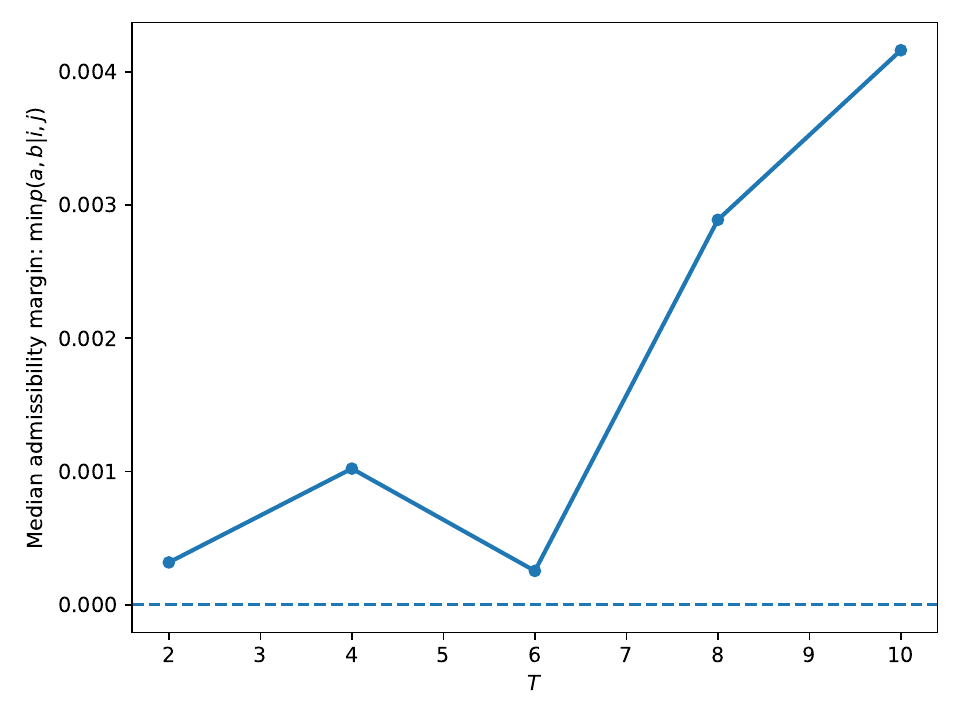}
\caption{Admissibility margin across walk times.
For each $T$, the plotted value summarizes (over seeds) the typical minimum joint probability
$\min_{a,b,i,j} p(a,b|i,j)$ associated with the best coarse-grained witnesses.
The finite-time violations occur while maintaining nontrivial admissibility margins, indicating that the observed post-quantum correlations remain operationally well-defined.}
\label{fig:A_minp_med_vs_T}
\end{figure}

The resulting maximal CHSH value $S_{\mathrm{coarse}}(T)$ is shown in
Fig.~\ref{fig:A_S_vs_T}, while Fig.~\ref{fig:A_frac_gt2} reports the fraction of admissible trials
yielding a violation of the classical bound $S>2$ across multiple random seeds.
Strikingly, we find a clear finite-time window in which coarse-grained position measurements
\emph{do} access some signature of the post-quantum correlations, even though the value of $S$ is inside the valid quantum window bellow $|S|\le2\sqrt2$~\cite{Tsirelson1980}.
In particular, for $T=4$ and $T=6$, the optimized witnesses robustly exceed the classical bound,
with typical values $S_{\mathrm{coarse}}\simeq2.1$ at $T=4$ and $S_{\mathrm{coarse}}\simeq2.03$ at
$T=6$.
By contrast, for $T\ge8$ the maximal observed value drops below $S=2$, and violations become
statistically suppressed.
This trend is reflected both in the seed-by-seed best values (Fig.~\ref{fig:A_S_vs_T})
and in the diminishing fraction of trials exhibiting $|S|>2$ (Fig.~\ref{fig:A_frac_gt2}). Fig.~\ref{fig:A_minp_med_vs_T} shows the admissibility margin across walk times, for comparison.

This behavior admits a clear physical interpretation.
At small $T$, the walker occupies only a few position sites, and the coarse-grained binning
retains sufficient resolution to probe the effective low-dimensional Schmidt subspace responsible
for Bell inequality violations.
As the walk time increases, the position Hilbert space grows linearly with $T$, while the
measurement resolution remains fixed.
The resulting mismatch progressively washes out the accessible correlations, even though the
underlying coin--position state continues to support large post-quantum violations when probed
with appropriately tailored measurements.

From an operational perspective, this finite-time window is particularly significant.
It demonstrates that post-quantum correlations are not merely a mathematical artifact of
high-dimensional measurements, but can emerge under simple, physically natural coarse-grained
observables, provided the system size is appropriately matched to the measurement resolution.
In experimental implementations of coined quantum walks—such as photonic, trapped-ion, or
superconducting platforms—the number of steps $T$ is directly tunable.
The present results therefore identify a concrete regime in which post-quantum correlations could
be operationally emulated using only standard quantum dynamics, standard projective measurements,
and classical post-processing, without requiring access to the full Schmidt-aligned measurement
structure.

Taken together with the large-$T$ benchmark results, this finite-time analysis reinforces the
central message of the paper: post-quantum correlations can arise in completely standard quantum
walk dynamics, while their operational accessibility depends sensitively on the interplay between
Hilbert-space growth and measurement resolution.

We emphasize that this analysis does not modify the Bell scenario itself:
only the walk time $T$ and the coarse-graining threshold are varied,
while the measurement structure and admissibility constraints remain unchanged.

We further performed independent coarse-grained scans and multi-start refinements (alternative grids and enlarged
parameter domains) to test robustness against local extrema. None yielded a larger admissible violation than the one
reported, supporting the stability of the estimate.

\section{Discussion and conclusion}
We have shown that a fully standard coined discrete-time quantum walk can generate admissible
post-quantum correlations in a coin--position Bell scenario, without any modification of its
underlying dynamics.
All post-quantum behavior arises exclusively from the extended preparation of the coin and from
the subsequent analysis of measurement statistics, while the walk evolution itself remains
unitary, nearest-neighbor, and entirely conventional. From an operational viewpoint, this interpretation follows directly from the complementarity-based
framework introduced in Ref.~\cite{deOliveira2020Complementarity}, in which physical consistency is
enforced at the level of observable statistics rather than state positivity.

This establishes quantum walks as a controlled setting in which post-quantum correlations can be
studied without invoking exotic dynamics or signaling mechanisms.
It is important to emphasize that we are not proposing a physically implementable quantum
procedure that prepares positive states violating Tsirelson’s bound.
Rather, our results identify operationally admissible post-quantum correlations that coexist with
fully standard quantum dynamics when positivity is relaxed at the level of preparation and
physical consistency is enforced solely through admissibility and no-signaling of observable
statistics.
As the extended coin operators considered in this work,
$\rho_c(\mathbf r)=\frac12(\mathbb I+\mathbf r\cdot\boldsymbol\sigma)$ with $\|\mathbf r\|>1$,
are not physical quantum states, they cannot be prepared directly within standard
quantum mechanics.
Nevertheless, their induced measurement statistics can be emulated operationally using a
quasiprobability decomposition into physical coin states followed by classical post-processing.

Indeed, for any Bloch vector $\mathbf r=\|\mathbf r\|\hat{\mathbf r}$ one may write
\begin{equation}
\rho_c(\mathbf r)
=\frac{1+\|\mathbf r\|}{2}\,\rho_{+\hat{\mathbf r}}
+\frac{1-\|\mathbf r\|}{2}\,\rho_{-\hat{\mathbf r}},
\label{eq:quasi_decomp}
\end{equation}
where $\rho_{\pm\hat{\mathbf r}}=\frac12(\mathbb I\pm\hat{\mathbf r}\cdot\boldsymbol\sigma)$ are
rank-one projectors onto antipodal pure coin states.
For $\|\mathbf r\|>1$, the second weight in Eq.~\eqref{eq:quasi_decomp} is negative, reflecting the
nonpositivity of $\rho_c(\mathbf r)$.

Operationally, Eq.~\eqref{eq:quasi_decomp} defines a quasiprobability protocol in which the coin is
prepared in one of the physical states $\rho_{\pm\hat{\mathbf r}}$, the standard quantum walk and
measurements are performed, and the observed frequencies are combined with signed classical
weights.
Expectation values of any observable $O$ are then reconstructed as
$\langle O\rangle_{\rho_c(\mathbf r)}
=\sum_{k=\pm} w_k \langle O\rangle_{\rho_{k\hat{\mathbf r}}}$,
with $w_\pm=(1\pm\|\mathbf r\|)/2$.
The corresponding quasiprobability \emph{negativity cost} can be quantified by the $\ell_1$ norm of the weights,
\begin{equation}
\mathcal{N} \equiv \sum_{k=\pm}|w_k| = \|r\|.
\end{equation}
Operationally, this implies an increased sampling overhead
in Monte-Carlo style reconstructions, with typical estimation
variance scaling as $\mathrm{Var}(\hat{E}) = O(\mathcal{N}^2/N_{\mathrm{shots}})$
for $N_{\mathrm{shots}}$ experimental shots.
The price of leaving the quantum state space is an increased sampling overhead quantified by
$\Nneg=\sum_k|w_k|=\|\mathbf r\|$, while no modification of the walk dynamics or measurement apparatus is required.

\paragraph{Experimental-ready emulation protocol.}
Equation~\eqref{eq:quasi_decomp} yields a direct, experimental procedure to emulate the extended
preparation without altering quantum dynamics.
For a chosen $\mathbf r=\|\mathbf r\|\hat{\mathbf r}$, one runs two standard experiments:
(i) prepare the coin in the physical pure state $\rho_{+\hat{\mathbf r}}$, implement the usual
$T$-step Hadamard walk, and estimate the four tables $p_{+}(a,b|i,j)$ from repeated runs;
(ii) repeat with $\rho_{-\hat{\mathbf r}}$ to obtain $p_{-}(a,b|i,j)$.
The target extended statistics are then reconstructed by the signed classical combination
$p(a,b|i,j)=w_{+}\,p_{+}(a,b|i,j)+w_{-}\,p_{-}(a,b|i,j)$ with weights
$w_{\pm}=(1\pm\|\mathbf r\|)/2$.
This protocol requires only standard state preparation, standard walk dynamics, and standard
coin/position measurements; the post-quantum features enter exclusively through the reconstruction rule.
The finite-time coarse-grained window identified in Sec.~IV further indicates that the required
measurement structure can remain experimentally simple when $T$ is chosen in the appropriate regime.

A central outcome of our analysis is the clear separation between the \emph{existence} of
post-quantum correlations and their \emph{operational accessibility}.
Schmidt-aligned observables reveal the maximal correlations supported by the walk-generated
coin--position state and allow violations beyond Tsirelson’s bound while respecting admissibility
and no-signaling.
In contrast, physically natural coarse-grained position measurements may fail to access the
relevant effective Schmidt subspace and thereby strongly suppress observable Bell violations.
This suppression is made explicit within a restricted but experimentally motivated family
of coarse-grained position binnings, and the walker profile shown in Fig.~\ref{fig:A_Px}
confirms that the effect is not tied to any nonstandard transport feature of the walk.
 Operationally enforcing physical consistency at the level of observable statistics, rather than
through a priori positivity constraints, has also appeared in different contexts, for example in
the construction of physically meaningful covariance matrices via conditioning procedures based on
Schur complements in continuous-variable systems~\cite{Haruna_2007}.

From an operational perspective, these results highlight that the absence of post-quantum
violations in a given Bell test does not necessarily reflect a fundamental limitation of the
underlying system, but may instead arise from restrictions imposed by the available measurement
structure.
More broadly, our work positions quantum walks as a versatile and transparent platform for
exploring the boundary between quantum and post-quantum correlations, emphasizing that this
boundary is shaped not only by dynamical constraints, but also by the interplay between state
preparation and measurement.
Post-quantum Bell behavior can be operationally realized
in a setting where the dynamics is strictly local and unitary, and where the only extension lies in
the preparation rule. Quantum walks are especially suited for this purpose because they convert a
minimal two-level coin into a rapidly expanding walker Hilbert space, enabling strong Bell
correlations to live in delocalized mode subspaces that are invisible to simple position
coarse-graining.

Although the post-quantum correlations identified in this work are not directly accessible under the coarse-grained measurement schemes considered, the observed separation between existence and accessibility highlights a fundamental constraint on the operational use of correlations in quantum communication protocols. In particular, it suggests that the performance of such protocols may be limited not only by the properties of the underlying quantum states, but also by the structure and resolution of the available measurements. This perspective opens the possibility that enhanced measurement strategies could unlock otherwise hidden nonclassical resources.

Future directions include the exploration of alternative coarse-grainings that better approximate
projectors onto the relevant Schmidt subspaces, extensions to higher-dimensional coins or
multi-walker settings, and experimental implementations of the underlying quantum-walk dynamics
in photonic or trapped-ion architectures.
From an experimental perspective, the walk dynamics considered here is already routinely realized
in several platforms, including photonic circuits, trapped ions, and superconducting qubits.
The central experimental challenge identified by our results lies not in the implementation of
the walk itself, but in the preparation and measurement structures required to access the
effective Schmidt subspaces that support post-quantum correlations within the extended operational
framework considered here.

While simple coarse-grained position measurements are naturally accessible, approximating
Schmidt-aligned observables would require more structured, possibly collective, measurements on
the walker degrees of freedom.
Our results therefore identify measurement structure, rather than dynamics, as the primary
experimental bottleneck in accessing post-quantum correlations.

\begin{acknowledgments}

M.C.O. acknowledges partial financial support from the National Institute of Science and Technology for Applied Quantum
Computing through CNPq (Process No.~408884/2024-0) and from the São Paulo Research Foundation (FAPESP), through the Center
for Research and Innovation on Smart and Quantum Materials (CRISQuaM, Process No.~ 2024/00998-6).
\end{acknowledgments}

\begin{appendix}
    
\section{Numerical methods and optimization strategies}
\label{sec:numerics}

All numerical results reported in this work are obtained by explicit simulation of the
coined discrete-time quantum walk and by direct evaluation of Bell correlators from the
resulting coin--position operator.
No approximations beyond finite numerical precision are employed.
In particular, the walk dynamics is implemented exactly as defined in
Sec.~III, and all Bell-test quantities are computed from the final operator $\rho_T$
using the Born rule.

The numerical analysis is organized around two complementary optimization strategies,
referred to as (Schmidt-aligned benchmark) and 
(coarse-grained operational search).
These two approaches address distinct conceptual questions and employ different numerical
procedures, which we summarize below.

\subsection{ Schmidt-aligned benchmark (existence of post-quantum correlations)}
\label{subsec:optionB}

This Schmidt-aligned benchmark is designed to characterize the \emph{maximal Bell correlations
supported by the walk-generated coin--position state}, independently of operational restrictions.
It therefore serves as a ceiling or benchmark for the correlations that can exist in
principle.

The procedure consists of two steps.

\paragraph{Quantum benchmark at $\|\mathbf r\|=1$.}
We first consider a standard quantum preparation with $\|\mathbf r\|=1$, for which the
initial coin operator corresponds to a pure state.
The walk evolution then produces a pure bipartite state
$\ket{\psi_T}\in\Hc\otimes\Hw$.
We reshape this state as a $2\times(2T+1)$ matrix and perform an exact Schmidt decomposition,
yielding two nonzero Schmidt coefficients and their associated coin and position Schmidt
vectors.

From the Schmidt coefficients we compute the theoretical maximal CHSH value using the
Horodecki criterion.
We then explicitly construct the corresponding optimal CHSH observables in the canonical
two-qubit form and map them back into the physical coin and position Hilbert spaces:
coin observables are rotated by the Schmidt basis on $\Hc$, while Bob’s observables are
embedded into the full position space by projecting onto the two-dimensional Schmidt
subspace and acting trivially on its orthogonal complement.
The numerical CHSH value obtained from these observables exactly saturates the theoretical
maximum, providing a stringent consistency check of both the walk simulation and the Bell
test.

\paragraph{Fixed-measurement scan for $\|\mathbf r\|>1$.}
Having fixed the Schmidt-aligned observables, we extend the coin preparation to
$\|\mathbf r\|>1$.
For each value of $\|\mathbf r\|$, we sample preparation directions
$\hat{\mathbf r}$ uniformly on the Bloch sphere and evolve the corresponding initial
operator $\rho_0=\ket{0}\!\bra{0}\otimes\rho_c(\mathbf r)$ under the same walk dynamics.

For each sample we compute the four joint probability tables $p(a,b|i,j)$ associated with
the fixed Schmidt-aligned measurements.
Candidates are retained only if they satisfy two operational constraints:
(i) admissibility, meaning that all probabilities are nonnegative and normalized, and
(ii) no-signaling at the level of observed marginals, within numerical tolerance.
Among all admissible candidates, we record the largest CHSH value $|S|$ and the minimum
joint probability as an admissibility margin.

This procedure produces the curves shown in
Figs.~\ref{fig:B_S_vs_r} and \ref{fig:B_minp_vs_r}, as well as the representative maximal
violation reported in Sec.~IV.A.
Importantly, no optimization over measurements is performed at $\|\mathbf r\|>1$:
the measurements are fixed once and for all by the Schmidt structure of the quantum
benchmark.
This ensures that the procedure isolates the effect of the extended preparation alone.

\subsection{Coarse-grained operational search (accessibility of post-quantum correlations)}
\label{subsec:optionA}

This choice addresses a different question, namely whether post-quantum correlations remain
\emph{operationally accessible} when Bob is restricted to simple, physically natural
coarse-grained position measurements.
In contrast to the previous one, the measurements are not tailored to the Schmidt structure of the
state and are instead fixed \emph{a priori}.

Bob’s observables are chosen as
$B_0(x)=\sgn(x)$ and a single-threshold binning
$B_1(x)=+1$ for $|x|\ge x_0$ and $-1$ otherwise.
The threshold position $x_0$ is expressed as a fraction of the walk time $T$ and is selected
based on a coarse preliminary scan.

For fixed values of $T$, $\|\mathbf r\|$, and $x_0/T$, the optimization proceeds via a
randomized search.
Each trial consists of:
(i) sampling a preparation direction $\hat{\mathbf r}$ uniformly on the sphere;
(ii) evolving the corresponding operator $\rho_T$ under the standard walk dynamics;
(iii) sampling two independent coin measurement directions
$\hat{\mathbf a}$ and $\hat{\mathbf a}'$ uniformly on the sphere;
(iv) computing the four joint probability tables $p(a,b|i,j)$ from $\rho_T$; and
(v) accepting the candidate only if admissibility and no-signaling constraints are satisfied.

For accepted candidates we compute the CHSH value and retain those yielding the largest
$|S|$.
This randomized procedure is repeated for many independent trials and preparation directions,
allowing us to identify the best admissible witness within the restricted measurement family.
The resulting witness parameters and correlators are reported in Table~\ref{tab:A_best},
and the corresponding walker position distribution is shown in Fig.~\ref{fig:A_Px}.

Unlike the Schmidt-aligned benchmark, this procedure does not aim to approximate the theoretical
maximum Bell violation.
Instead, it quantifies how much of the post-quantum behavior revealed by the Schmidt-aligned
benchmark survives under operationally simple and experimentally motivated measurement
restrictions.

\subsection{Numerical stability and tolerances}

All admissibility and no-signaling checks are enforced with explicit numerical tolerances,
typically at the level of $10^{-7}$--$10^{-9}$.
We have verified that tightening these tolerances does not qualitatively affect any of the
reported results.
The qualitative conclusions---existence of admissible post-quantum correlations in the first choice
and their strong suppression under coarse-graining in the second choice---are robust against changes in
random seeds, sampling density, and walk time $T$.

\end{appendix}

\bibliography{refs}

\end{document}